\documentclass[useAMS,usenatbib]{mn2e}
\usepackage{graphicx}
\usepackage {subfigure}
\usepackage[authoryear]{natbib}
\usepackage{rotating}
\newcommand{\phr}{Pegase-HR}
\newcommand{\vaz}{Vazdekis/Miles}

\title[Ages and metallicities from full spectrum fitting.]{Spectroscopic ages 
and metallicities of stellar populations: validation of full spectrum fitting}

\author[M. Koleva et al.]{M. Koleva$^{1,2}$\thanks{E-mail:
mina.koleva@obs.univ-lyon1.fr}, Ph. Prugniel $^{1, 3}$, P. Ocvirk $^{4,5}$, D. Le Borgne $^{4}$
  and C. Soubiran  $^{6}$\\
$^{1}$Universit\'e de Lyon, Lyon, F-69000, France ; Universit\'e Lyon~1,
Villeurbanne, F-69622, France; Centre de Recherche Astronomique de
Lyon, \\ 
Observatoire de Lyon, St. Genis Laval, F-69561, France ; CNRS, UMR 5574 ; \\
$^{2}$Department of Astronomy, St. Kliment Ohridski University of Sofia, 5 James
Bourchier Blvd., BG-1164 Sofia, Bulgaria \\
$^{3}$GEPI, Observatoire de Paris, 61, avenue de l'Observatoire, F-75014 Paris\\
$^{4}$Laboratoire AIM, CEA/DSM - CNRS - Universit\'e Paris Diderot, DAPNIA/Service d'Astrophysique, CEA-Saclay, 91191 Gif-sur-Yvette, France\\
$^{5}$Astrophysikalisches Institut Potsdam, An der Sternwarte 16, D-14482 Potsdam, Germany\\
$^{6}$Observatoire Aquitain des Sciences de l'Univers, 2 rue de l'Observatoire, 33270, Floirac, France; CNRS, UMR 5804}
\begin{document}

\date{Accepted 2007 December 31.  Received 2007 December 22; in original form 2007 August 15}

\pagerange{\pageref{firstpage}--\pageref{lastpage}} \pubyear{2008}

\maketitle

\label{firstpage}

\begin{abstract}

Fitting whole spectra at intermediate spectral resolution (R = 1000 -- 3000),
to derive physical properties of stellar populations, appears
as an optimized alternative to methods based on spectrophotometric indices:
it uses all the redundant information contained in the signal.
This paper addresses the validation of the method and it investigates the
quality of the population models together with the reliability of the fitting
procedures.
Our method compares observed optical spectra with models to derive the 
age, metallicity, and line broadening due to the internal kinematics. It
is insensitive to the shape of the continuum and
the results are consistent with Lick indices but three times more precise.
We are using two algorithms: STECKMAP, 
a non-parametric regularized program and NBURSTS a parametric non-linear
minimization.

We compare three spectral synthesis models for single stellar populations: \phr, Galaxev (BC03)
and \vaz, and we analyse spectra of Galactic clusters
whose populations are known from studies of color-magnitude diagrams (CMD)
and spectroscopy of individual stars.

We find that: (1) The quality of the models critically depends on the
stellar library they use, and in particular on its coverage in age, metallicity and 
surface gravity of the stars. \phr\ and \vaz\ are consistent, 
while the comparison between \phr\ and BC03 shows some
systematics reflecting the limitations of the stellar library
(STELIB) used to generate the latter models; 
(2) The two fitting programs are consistent;
(3) For globular clusters and M67 spectra, the method restitutes 
metallicities in agreement with spectroscopy of stars with a precision of 0.14~dex;
(4) The spectroscopic ages are very sensitive to the presence of
a blue horizontal branch (BHB) or of blue stragglers. A BHB morphology
results in a young SSP-equivalent age. Fitting a free amount of blue stars
in addition to the SSP model to mimic the BHB
 improves and stabilizes the fit and restores ages in agreement with CMDs studies.
 This method
is potentially able to disentangle age or BHB effects in extragalactic
clusters. Full spectrum fitting is a reliable method
to derive the parameters of a stellar population.

\end{abstract}


\section{Introduction}

The goal of this paper is to validate a method for determining the
characteristics of a stellar population using an entire medium resolution
spectrum in the rest-frame optical range (rather than concentrating on specific 
features). Special attention is given to make an optimal use of the information, 
to look for systematics and to compare the different models of stellar populations.

Stellar populations keep a record of most of the evolution of the stellar
systems, and by studying Hertzprung-Russell (HR) diagrams 
(or colour-magnitude diagrams, CMDs) of  galaxies, 
it is possible to constrain  
the history of the star formation and enrichment since their formation 
\citep[see eg.][]{grebel05}. 
Unfortunately, these star-resolved observations are restricted to
nearby galaxies (Local Group and a little further) because of the 
shallow magnitude limits and crowding effect. At large distances, only 
the spectral energy distributions (SED) 
and the line-of-sight (LOS) integrated spectra are available.

Many methods were devised to determine the age and the metallicity
of stellar systems from their integrated light. 
In this paper we distinguish three approaches:
\begin{enumerate}
\item SED fitting  - sensitive
to the general shape of the continuum, (usually) made in broad wavelength 
ranges;
\item Spectrophotometric indices -
concentrates on the strength or equivalent width of specific features,
not sensitive to the flux calibration;
\item Full spectrum fitting - use all the information by fitting all the
pixels, but independently from the shape of the continuum;
\end{enumerate}

 SED fitting  \citep{cid05,pan,sol05}  can be made using colours,
 low or medium resolution spectra. It is performed
 over large wavelength ranges and
compares the observations to models like {\sc P\'egase}.2 
\citep{pegase,pegase2},~{\sc Galaxev} \citep{BC03}
or others. These models should take into account the various sources 
of extinction (internal and Galactic). SED fittig constrains not only
the stellar population but it also informs about the dust content.

Using low- and medium resolution spectra, the classical approach 
is to measure spectrophotometric indices, like Lick 
\citep{faber73,wor94} or \citet{rose84}
indices and compare them to model predictions. 
Spectrophotometric indices measure the strength of some spectral features 
expected to trace the abundance of particular elements or to correlate with 
the mean age of stellar populations.
However, primarily because a galaxy spectrum is naturally broadened by
the internal kinematics, the spectrophotometric indices are defined at low 
resolution (a few {\AA}) and are therefore blends of many spectral lines.
That is why their responses to age and metallicity are quite complex and
require careful calibrations \citep{K05}.
Attempts to define indices at a higher resolution \citep[see][]{PHR}
require delicate
velocity-dispersion corrections or must be designed carefully to reduce
their sensitivity to the velocity dispersion. A good example
of high-resolution index is the H$\gamma$ index defined in \citet{va99}.
Another possibility to improve the sensitivity of an index to either
age or metallicity is to use a composite definition, like for the 
CaT$*$ (calcium triplet corrected for Paschen lines, \citealt{cenarro01}) 
or MgFe (insensitive to [$\alpha$/Fe], \citealt{G93,tmb03}) 
indices.

Lick indices are widely used for more than a decade and the methods
continue to improve with introducing the effects of non-solar abundances.
However, they
use only a few small passbands out of the full observed spectral range,
and although the information in the spectrum is probably redundant, 
this does not optimize the usage of the collected light.
Another limitation of the indices is that they are sensitive to
missing data or bad pixels \citep{Chil07}: The interpolation
of spectral bins affected by detector defects or cosmic hits may bias the 
measurement of the index if the observations are close to the sampling limit. 
The indices are also affected by weak emission lines \citep{GE96}.

The full spectrum fitting that we propose 
combines the advantages of spectrophotometric indices and SED fitting: 
It uses all the information, it is insensitive to extinction or
flux calibrating errors and it is not limited by the physical broadening,
since the internal kinematics is determined simultaneously with the population
parameters. 
>From the point of view of
measuring the internal kinematics, it is an extension of the classical
methods (eg. optimal template fitting, \citealt{vdmarel94}).  The method
has been presented in several places \citep{pru2003,ocv06a,ocv06b} 
and its reliability assessed on the basis of various simulations.
It is three times more sensitive, but still consistent with 
Lick indices \citep{Chil07,mich07}.

Full spectrum fitting is well suited when the resolution 
is comparable with the physical broadening 
(10~km\,s$^{-1} <  \sigma_{ins} < $ 200~km\,s$^{-1}$ ). 
At lower resolution (R = 500 -- 1000) the spectral features are diluted
and the sensitivity of the method would rely on the strong spectral
features and therefore would become equivalent to Lick indices.
At even lower resolution, 
only strong features, like the Balmer break provides information on the
stellar population. In this case, SED fitting using also the shape of the continuum 
is the good solution.

In its intention to be insensitive to the flux calibration and extinction
our full spectrum fitting is close to spectrophotometric indices, but it uses
all the spectrum. A possible disavantage compared to Lick indices is that the
 method may, in principle, be
sensitive to the wavelength range and to spectral resolution. However, 
\citet{kol06} have shown that this sensitivity is not critical:
the precision of the results mostly depends on the total S/N (i.e.
integrated over all the spectral range), and even excluding the Balmer lines
does not bias the age determination  (only precision is reduced).

To check the reliability of a spectroscopically determined
SSP equivalent age/metallicity, it is wise to
 test it on globular clusters
(GC) and to compare the results with the results from CMDs determinations. 
It has been shown \citep{wolf07} that the
full spectrum fitting 
reproduces better the results from the CMDs than the other methods.

This paper is organised as follows:
In Sect. 2 we present the population models and the two algorithms
implementing the full spectrum fitting. 
In Sect. 3, we use these algorithms to compare different
synthetic population models and in Sect. 4 we invert spectra of Galactic
clusters and compare the results with age determinations from HR diagrams.
Section 5 discusses the prospects for analyzing stellar
populations and gives conclusions.

\section{Population models and fitting method}

The goal of this analysis is to retrieve simultaneously the characteristics
of the internal kinematics and of the stellar population using a
medium resolution spectrum that 
integrates along the line-of-sight the light from a region of a galaxy 
or cluster (hereafter LOS-integrated spectrum).

The main characteristics of this method are (1) to be  insensitive to the
shape of the continuum and (2) use all the information in the spectrum.
Its principle is to compare an observation to a
population model convolved with a line-of-sight velocity distribution
(LOSVD).
The spectrum of the population model must have the same
line-spread function (LSF) as the observation (i.e. the LSF of the observation
must be injected into the model).

In this section we describe the main ingredients of the method used to analyse 
stellar populations. First we present the population models and the stellar 
libraries used to construct the synthetic spectra.
Then we describe the two algorithms used to retrieve the stellar population 
kinematics and characteristics from the spectra.

\subsection{Spectral synthesis models for stellar populations}

The population model, i.e. the spectrum corresponding to some formation 
scenario including a given evolution, is constructed using a population synthesis
code. There exists several such codes, and in this paper we are using
\phr\ \citep{PHR}, Galaxev (BC03)
and \vaz\
(Vazdekis et al. in preparation).
These codes differ
by their choice of the physical ingredients (IMF, evolutionary tracks, ...),
by the libraries of stellar spectra they are using and by implementation
details (in particular interpolations).

\subsubsection{Pegase-HR and ELODIE 3.1}

The \phr\ code\footnote{http://www2.iap.fr/pegase/pegasehr/}, 
as described in \citet{PHR}, allows a choice of different IMF
and different physical ingredients.
The evolutionary tracks are from Padova group 
(Bressan et al. 1993 and companion papers: Padova~94). 
 They are identical
to those used for Pegase.2 \citep{pegase2}.
For the present work we use SSPs computed with Salpeter IMF
(with lower mass 0.1 $M_\odot$ and upper mass 120 $M_\odot$).

At variance with \citet{PHR}, we are using a new version of the 
ELODIE library\footnote{http://www.obs.u-bordeaux1.fr/m2a/soubiran/elodie\_library.html}
(version 3.1; \citealt{elo3.1}) at a resolution FWHM = $0.55$~{\AA} or
$R=\Delta\lambda/\lambda=10000$ at $\lambda = 5500$~\AA.  
The ELODIE library is constructed from the spectroscopic archive of
Observatoire de Haute-Provence\footnote{http://atlas.obs-hp.fr/elodie} 
\citep{moultaka04} and the principle of the data-reduction,
flux calibration and determination of the atmospheric parameters are
described in \citet{elo1}. The set of 1962 spectra of 1388 stars
is unchanged in the current release with respect to \citet{elo3}.
A comparison between the different theoretical and empirical libraries
made by \citet{martins07} shows that the coverage of the library
in effective temperature, surface gravity and metallicity
($T_{eff}$, $\log \mathrm{g}$ and [Fe/H]) is one of the best from the present
libraries and therefore well suited for population studies
(still, enlarging the library toward cooler stars and lower metallicity
is desirable, and is currently under preparation).

\begin{figure}
\includegraphics[width=0.48\textwidth]{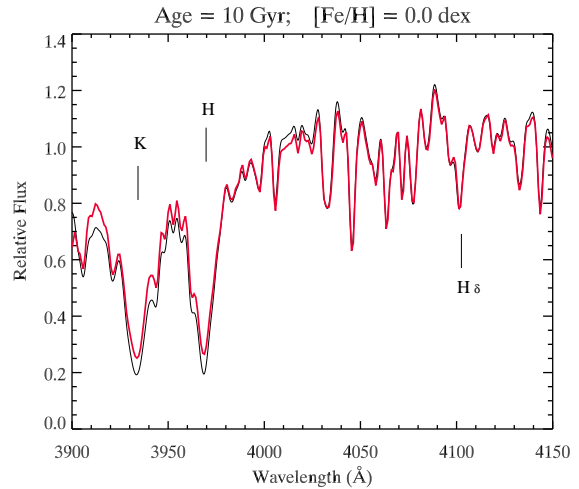}
\caption{Direct comparison between \phr/ELODIE.3.1 (in black) and 
\vaz\ (in red, thick line) 
populations of 10 Gyr, solar metallicity in the region of H \& K.
The \phr\ spectrum
was convolved by a LSF of 60~$km\,s^{-1}$ dispersion to match it to the
resolution of MILES, the vertical scale was adjusted to have
the same flux at 4000~{\AA}.
}
\label{fig:compHK}
\end{figure}

In the current release, the wavelength range, $3900 - 6800$~{\AA}
has been extended in the blue and includes the H \& K lines (Fig.~\ref{fig:compHK}).

In addition, the data processing and the determination of the atmospheric 
parameters have been also improved in several ways.
In particular, it was found that the diffuse light was under-subtracted by 
the standard reduction pipeline. 
This light is normally modeled using the inter-order
regions, but at the edge of the CCD where the first blue orders are located
the diffuse light displays steep variations and is not properly fitted by
the model. The under-subtraction of the diffuse light affects the blue
region of the spectrum, below about $4400$~{\AA}. We applied an ad'hoc
correction of this effect in order to avoid the full re-processing
from the initial CCD images (see \citealt{ersp02} for a fine tuned full
re-processing). 
The blue range, up to about $4400$~{\AA}
is affected by this small correction.

The comparison of a \phr\ SSP with Vazdekis/Miles, Fig.~\ref{fig:compHK},
shows that the models are visually consistent even in the
blue extremity where ELODIE spectra have lower S/N.
It also reveals that the flux calibration of ELODIE is losing
precision below $3970$~{\AA}. Though it is not critical for our
full spectrum fitting method, it may be of importance
for other applications. This degradation is due to the sensitivity
drop in the blue and to edge effect in the calibration process.

The isochrones used in \phr\ are scaled-solar at different values of the
total metallicity Z and this has two consequences on the modeling and
analysis of stellar populations:
\begin{enumerate}
\item It is well established and understood
that the metallic mixture of the Sun is not universal. For example, 
non-solar abundances of the $\alpha$-elements are surely the explanation
of the miss-fit of a strong spectral feature such as the Mg triplet and MgH
band near $5170$~{\AA} in elliptical galaxies \citep{worthey92}. The 
globular clusters are other important environment where the 
 $\alpha$-elements are enhanced, [Mg/Fe] $> 0$. 
Therefore, scaled-solar models are not strictly appropriate in those cases,
because the location of the isochrones in the HR diagram
depends on the detailed chemical composition
\citep{salaris98,kim02,salasnich00}.
Age determinations may be biased by up to $20$ per cent.
At low metallicity ([Fe/H] $< -0.7$~dex) the location of the isochrones 
mostly depends on the total metallicity, Z, while at higher metallicity 
the position of the giant branch is rather controlled by [Fe/H] 
(but a simple rescaling is not possible). This limitation will be
solved only when evolutionary tracks at non solar [$\alpha$/Fe] will be
implemented in \phr;

\item The stellar library is constituted from stars of the
solar neighborhood, and though detailed measurements of [$\alpha$/Fe]
are available for only a small fraction of the stars, we can assume
that it has the typical abundance pattern of this population (see e. g.
\citealt{wheeler89}; [$\alpha$/Fe] increases when [Fe/H] decreases).
 In \phr\ we neglected the change
of [$\alpha$/Fe] and of Z when [Fe/H] decreases (the predicted spectra
have the correct Fe content but have an excess of $\alpha$-elements at low
metallicity). This has to be considered when interpreting the metallicities
determined using these models;
\end{enumerate}

\begin{figure}
    \includegraphics[width=0.48\textwidth]{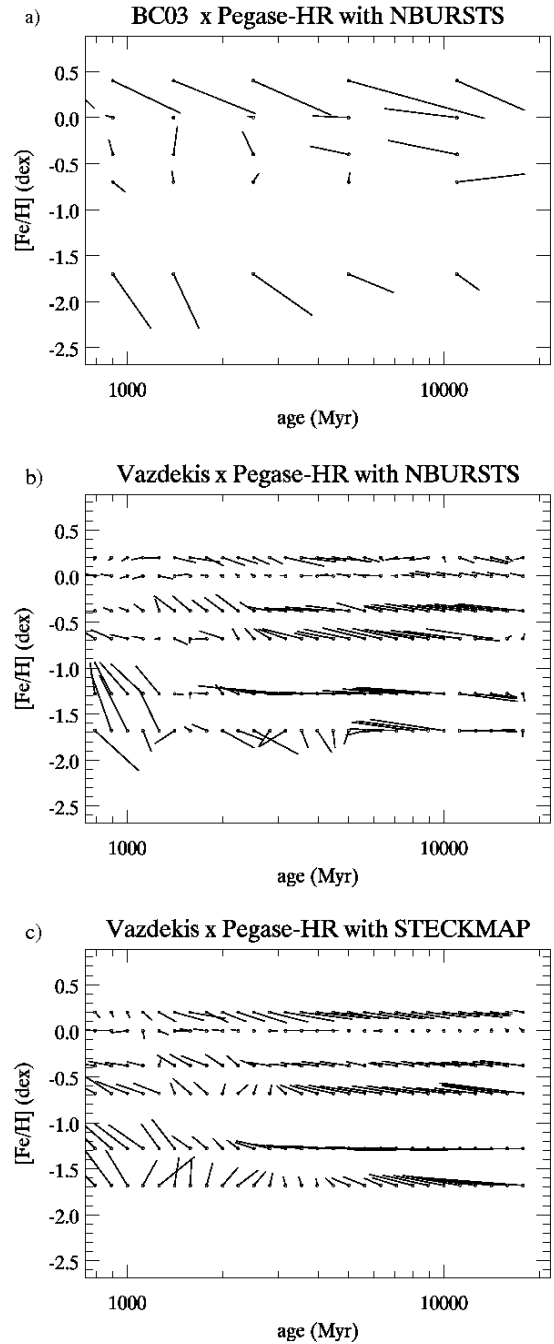}
    \caption{Fits of BC03 with NBURSTS (a),\vaz\ with NBURSTS (b) and
      \vaz\ with STECKMAP (c) SSPs, using \phr\ as a reference.
      Each vector represents an individual inversion.
      The dots are the location of the nominal
      characteristics of the population and the tip of the vectors are
      the values returned by the analysis procedure.
}
\label{fig:1}
\end{figure}

\subsubsection{Galaxev and the STELIB library}
The Galaxev models (\citealt{BC03}; commonly called BC03) use a variety of
evolutionary tracks: Padova~94 (as \phr), Padova 2000
\citep{girardi00} and Geneva \citep{geneva}.  After comparing these
different prescriptions, the authors adopted as 'standard' model the
Padova~94 set, because Padova~2000 would predict exagerately old
ages for elliptical galaxies (the red giant branch is $50-200$ K warmer
in Padova~2000).

\begin{figure}
\centering
\includegraphics[width=0.49\textwidth]{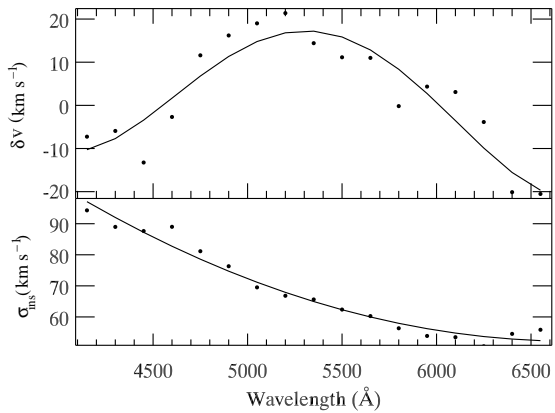}
\hspace{20pt}
\caption{Error of velocity and variation on instrumental velocity dispersion
 in BC03/STELIB models 
as a function of the wavelength. The dots represent the measured values and
 the lines are the polynomial fits with degrees of 4 and 2 for velocity and for dispersion, 
respectively.
}
\label{fig:bc_lsf}
\end{figure}

These models are build using the STELIB library \citep{stelib}, with a
resolution $\approx$ 3~{\AA}~FWHM in the spectral range $3200 - 9500$~{\AA}.
STELIB contains 249 spectra but only 187 stars have measured metallicity
 and can be used
to compute the predicted spectra. The coverage in the parameters space is
much more limited that in the other recent libraries \citep[see][]{martins07}.

\subsubsection{Vazdekis and the MILES library}
\vaz\ models\footnote{
http://www.ucm.es/info/Astrof/miles/miles.html} are based on the previous models from 
\citep{vaz99,vaz03}. 
They use Padova 2000 isochrones, which as we pointed before
has hotter red giant branch. 
They cover ages between
 0.1 and 17.5~Gyr and metallicities between -1.7 and 0.2~dex. 
The MILES library \cite{miles2006} used for these models have a 2.3~{\AA} spectral
resolution (FWHM) and cover the wavelength range from
3525 to 7500~{\AA}.
This library is believed to be better
flux calibrated than any other.

\subsection{Fitting methods}

\subsubsection{Pre-treatment: Line spread function matching}

In order to use SSP models to analyse observed spectra, a prior requirement is 
to match the spectral resolution of the models to that of the observations.

The spectral resolution is coarsely characterized by its FWHM, but entering
in the details, the instrumental broadening, or line-spread function 
(LSF, the spectral equivalent of the PSF for images), is not necessarily
a gaussian and varies with the wavelength and position in the field
(for integral field or long slit spectrographs). 
Therefore, matching the resolution is not a mere convolution.

Note that we are considering here only the instrumental broadening.
Observed spectral features are also broadened by the internal kinematics of 
the galaxies or clusters, and this {\it physical} broadening is 
measured by our analysis method.

\begin{figure*}
  \includegraphics[width=1\textwidth,height=0.9\textheight]{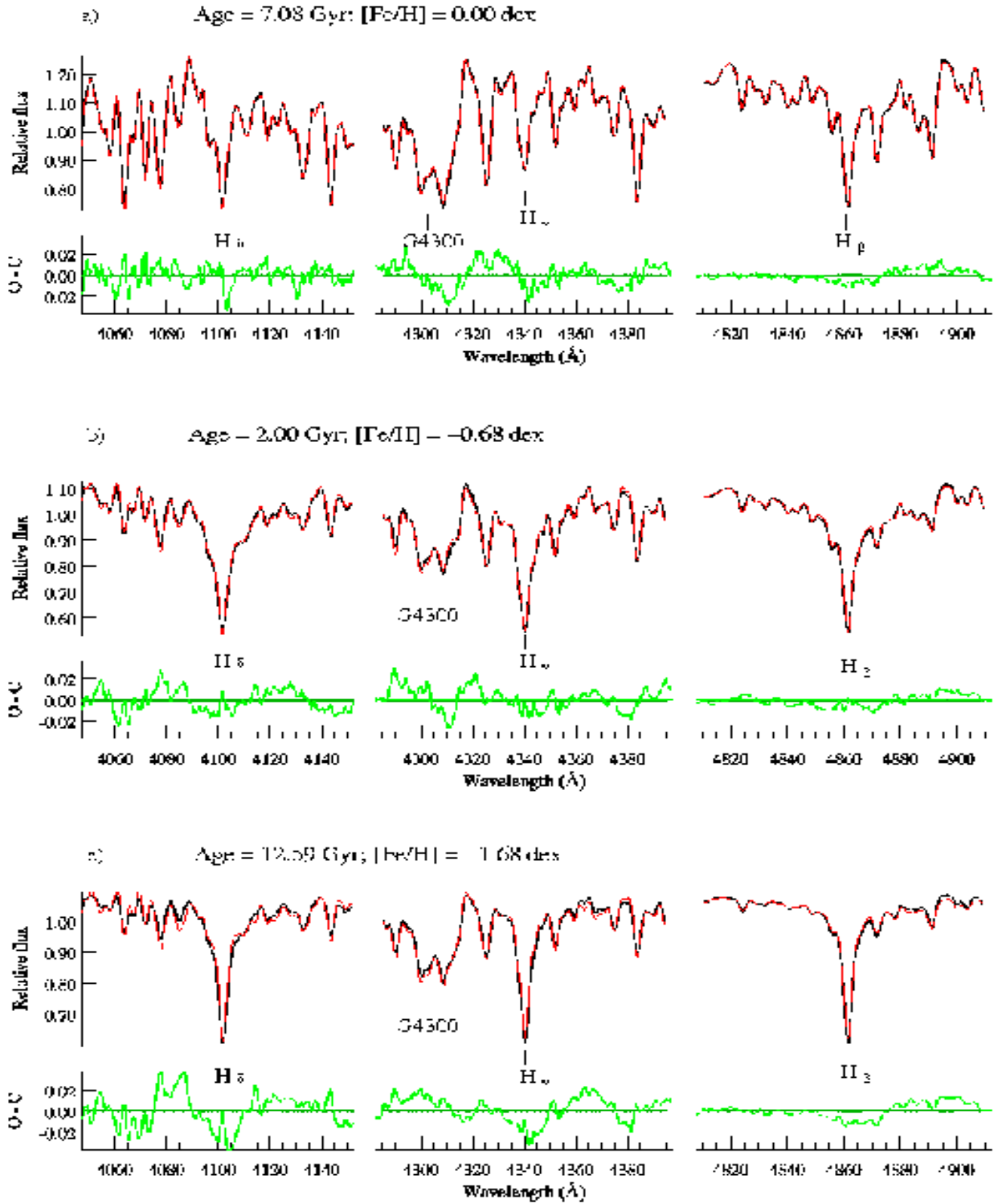}
    \hspace{10pt}
    \caption{
      Direct comparison between SSPs spectra from \vaz\
      (black line) and the SSPs of
      \phr/ELODIE.3.1 (red, thick line). 
      The ages/metallicities of the SSPs are:
      (a) 7~Gyr/0.0~dex; (b) 2~Gyr/$-0.68$~dex; (c) 12.6~Gyr/$-1.68$~dex.
      The \phr\ spectrum
      was convolved with a LSF of 60~$km\,s^{-1}$ dispersion to match it to the
      resolution of MILES. The spectra are normalized to 1.
      In green (O-C) we show the residuals.
    }
  \label{fig:vazphr}
\end{figure*}

The operation proceeds in two steps: First we determine the LSF
and its variation with the wavelength, and then we inject it
in the model. To be more precise, we do not need to determine the
intrinsic, but simply the relative LSF, i.e. the relative broadening 
function between the spectrum to analyse and the model.

In principle, when dealing with original observations, the LSF can be
determined using some template stars 
or using twilight spectra (i.e. solar spectrum). But in the present paper 
where we analyse models and stars clusters from a
library, we cannot access such calibrations. Therefore, we measure the
total broadening (instrumental plus physical). 

To determine the relative LSF we first used the program NBURSTS (see below)
to identify the best matching \phr\ SSP, and then we measured locally 
the LSF in 10 overlapping segments of the spectrum with the PPXF 
program\footnote{http://www.strw.leidenuniv.nl/\~{}mcappell/idl/} 
\citep{ppxf}. 

The ability to recover kinematical information distinguishes our
method from others which either fix the systemic velocity ($v_{sys}$)
and the velocity dispersion ($\sigma$)
or adopt
values from independent measurements. 
In the present case where
we analyse models and clusters, with no or very low physical velocity
dispersion,
we decreased the measured LSF to simulate a physical dispersion of 
 30~km\,s$^{-1}$.

Finally, the population models are convolved by all the local LSFs
and interpolated between these convolved segments 
to reproduce the wavelength variation of the real
LSF. The importance of the LSF injection cannot easily be judged from the
visual inspection
of the residuals between the 'observation' and the best-fit model, since
the fit
without this correction appears already excellent. But the effect 
could be seen
on the $\chi^2$ maps: the well in the minimum region becomes sharper
with the
LSF injection and the estimated error bars 
decrease.

To compute the values of the $\chi^2$ we are using the following formula:
$\chi^2 = [(observation-model)^2/errors^2]/n_{free}$,
where $n_{free}$ is the degree of freedom
and the errors are sum between the errors in the models
and the errors in the observations ($errors_{models}^2+errors_{obs}^2$).
The errors of the observations of globular clusters are taken from 
the library of \citealt{schia}.
For the errors of the models we adopted S/N=100 for \vaz\ and BC03
 and 150 for \phr\ 
(from visual inspection of residuals from the fits). These choices
affect only the $\chi^2$ scale
which should therefore be considered as relative, as values
lower than one on some of the maps reported in this paper are demonstrating.
Finally to produce $\chi^2$ maps we scan the space of age and metallicity,
fitting only the coefficients of LOSVD and multiplicative polynomial and
fixing the parameters of the SSPs (t,[Fe/H]).

\subsubsection{NBURSTS}

The program NBURSTS (see a preliminary description in \citealt{nbst}) 
performs a parametric  non-linear fit. 

The model used in this paper is a SSP convolved with a gaussian LOSVD and
multiplied by a polynomial whose 
degree is chosen to absorb any flux calibration systematics or effects
of extinction:
\begin{eqnarray}
{\rm{Obs}(\lambda)} = P_{n}(\lambda)\times(&SSP(t,[Fe/H],\lambda)\earth {}
 \nonumber\\
&{}LOSVD(v_{sys},\sigma, h3, h4)) \, 
\end{eqnarray}
\begin{figure}
  \centering
  \includegraphics[width=0.49\textwidth,height=0.9\textheight]{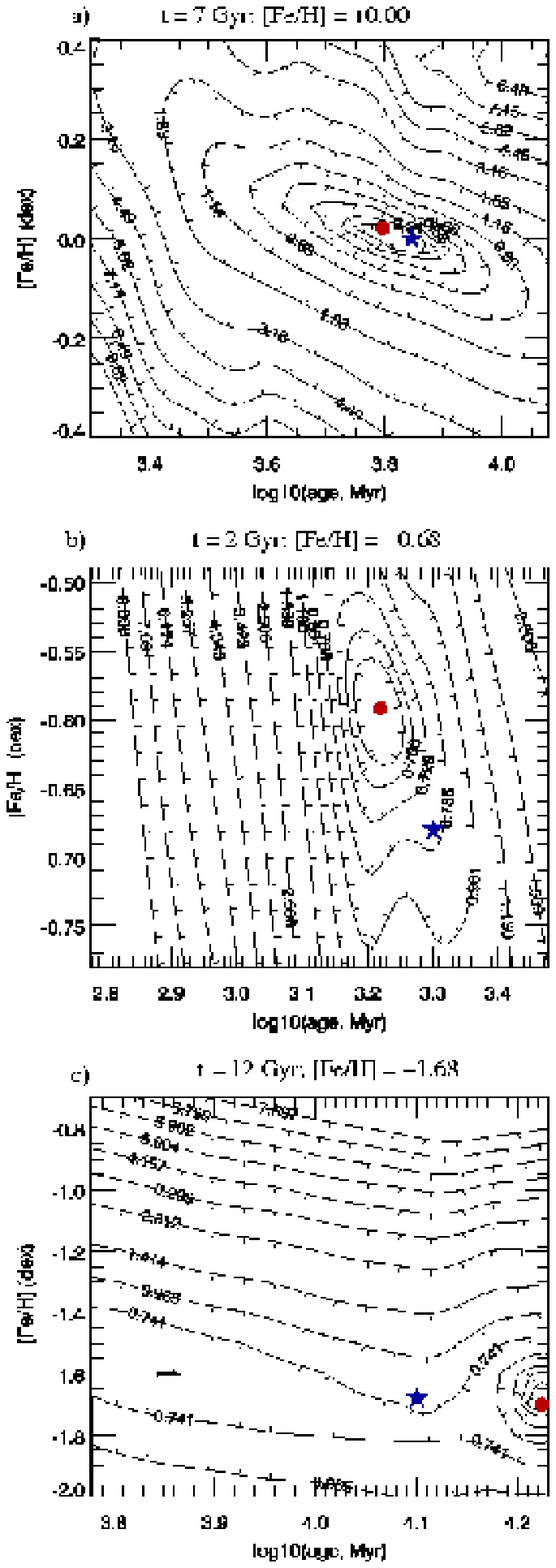}
  \caption{$\chi^2$ maps for the comparison between \vaz\ and \phr\ SSPs spectra of Fig.~\ref{fig:vazphr} ((a) [Fe/H] = 0.0~dex 
 $t=7$~Gyr; (b) [Fe/H] = $-0.68$~dex and 
    $t=2$~Gyr; (c) [Fe/H] = $-1.68$~dex, $t=12.59$~Gyr) inverted with \phr/ELODIE.3.1.
    The minimum is shown with a red circle. The blue star correspond to the age/metallicity
of the input SSP. The contours represent the $\chi^2$ iso-lines.
  }
  \label{fig:chi2vaz}
\end{figure}
The free parameters of the minimization are the characteristics of the SSP - 
age ($t$) and metallicity ($[Fe/H]$); the characteristics 
of the LOSVD ($v_{sys}$, $\sigma$, $h3$, $h4$), and the
$n+1$ coefficients of the multiplicative function $P_n$, a linear
combination of Legendre polynomials.

NBURSTS makes a Levenberg-Marquart minimization and can fit a combination
of SSPs or complex stellar population (in this paper we are fitting only SSPs). 
The population models are spline interpolated over a t-[Fe/H] grid of models
(a linear interpolation would not work because the derivatives must be
continuous).
It is written in IDL/GDL starting from PPXF and uses the 
MPFIT\footnote{Markwardt, http://cow.physics.wisc.edu/\~{}craigm/idl/idl.html} procedure. The fitting algorithm appears stable, because the free parameters are
independent enough, and no additional regularization was required.
At low signal-to-noise the high order terms of the LOSVD can be
neglected (e. g. fixed to 0).

The degree of the polynomial can be quite large. Experimenting 
with fits against \phr\ models (wavelength range $3900-6800$~{\AA}) we found
that the results were stable for degrees between 12 and 50. 
With $n = 50$ and a wavelength range of $3000$~{\AA}, the fit can 
absorb variations in the continuum at the scale of about $60$~{\AA}, which is 
comparable to Lick indices (linearly interpolating the continuum
between two side bands). 
The high order 
terms of the multiplicative polynomial are well decoupled from all 
the kinematical and population parameters. The $\chi^2$ maps with
$n = 50$ are more regular and have a sharper and deeper minimum. 
When two local minima exist, it is easier to identify the absolute one.
In principle, at an even higher degree the polynomial should become degenerated 
with the velocity dispersion and metallicity, but we did not find such an effect
(we did not explore degrees higher than 100).
The main inconvenient of a high degree is the
sensitivity to spikes of emission lines, which should be preliminary 
masked or clipped using a fit with a low degree polynomial.

It is important to stress that the inclusion of a multiplicative
continuum differs from the
'rectification' of the flux calibration used in some other algorithms,
like \citet{mathis06}, where the spectrum and the  model are divided by their 
top-hat filtered versions (width of 500 {\AA}).
Since both the observation and the model are similarly filtered the
results should not be biased but the sensitivity to the broad molecular 
bands of MgH, TiO or CN is probably reduced.
At variance, our approach does not affect the real spectral features, it is
rather similar to the method of Simien \& Prugniel (like in \citealt{sp02})
that renormalises the observations using a polynomial
fitted on the ratio between the observation and the best model and
iterate until convergence.

\begin{table*}
\centering
 \begin{minipage}{180mm}
\caption{Comparison of the ages/metallicities obtained with our method and those from the literature. Col. 1: ID of the spectra in \citet{schia} library consisting in (i) name of the cluster; (ii) (a,b,c) for observations in different nights; (iii)(1,2) for extraction in different apertures. Col. 2: The ages, metallicities and corresponding errors from full spectrum fitting with \phr; Col. 3: like Col. 2 but fitting also a fraction of hot stars. Col. 4: ages/metallicities/HBR from the literature. The HBR = (B-R)/(B+V+R), is taken from the Harris catalog \citep{harris96}}.
\begin{tabular}{@{}lrcrcrrr@{}}
\hline
& \multicolumn{2}{c}{Pegase-HR}&\multicolumn{2}{c}{Pegase-HR+hot stars}&\multicolumn{3}{c}{literature}\\
Name  &age [Gyr]& [Fe/H] &age [Gyr]& [Fe/H] & age [Gyr] & [Fe/H] & HBR \\
\hline
NGC0104 a 1 & 10.35 $\pm$  0.31 & -0.762 $\pm$ 0.006 & 11.26 $\pm$  0.38 & -0.747 $\pm$ 0.006 & 10.7 & -0.70$^a$ & -0.99 \\
NGC1851 a 1 &  5.32 $\pm$  0.11 & -1.163 $\pm$ 0.007 &  7.64 $\pm$  0.47 & -1.131 $\pm$ 0.009 &  9.2 & -1.21$^c$ & 
-0.36 \\
NGC1904 a 1 & 11.99 $\pm$  0.57 & -1.948 $\pm$ 0.012 & 14.02 $\pm$  0.25 & -1.884 $\pm$ 0.021 & 11.7 & -1.55$^c$ & 
-0.89 \\
NGC1904 b 1 & 14.94 $\pm$  0.14 & -1.891 $\pm$ 0.011 & 14.06 $\pm$  0.19 & -1.878 $\pm$ 0.015 &  &  \\
NGC2298 a 1 & 14.44 $\pm$  0.42 & -2.033 $\pm$ 0.024 & 13.84 $\pm$  0.56 & -1.961 $\pm$ 0.033 & 12.6 & -1.97$^a$ & 
 0.93 \\
NGC2298 b 1 & 10.25 $\pm$  0.63 & -2.123 $\pm$ 0.018 & 11.04 $\pm$  0.85 & -1.999 $\pm$ 0.027 &  &  \\
NGC2808 a 1 &  6.77 $\pm$  0.19 & -1.155 $\pm$ 0.010 &  8.00 $\pm$  0.41 & -1.105 $\pm$ 0.007 & 10.2 & -1.29$^c$ & 
-0.49 \\
NGC2808 b 1 &  6.92 $\pm$  0.21 & -1.161 $\pm$ 0.011 &  7.97 $\pm$  0.32 & -1.104 $\pm$ 0.008 &  &  \\
NGC3201 a 1 &  5.10 $\pm$  0.16 & -1.232 $\pm$ 0.010 &  7.68 $\pm$  0.78 & -1.195 $\pm$ 0.013 & 11.3 & -1.56$^a$ & 
 0.08 \\
NGC3201 b 1 &  7.04 $\pm$  0.31 & -1.351 $\pm$ 0.015 & 10.87 $\pm$  1.19 & -1.308 $\pm$ 0.015 &  &  \\
NGC5286 a 1 & 14.20 $\pm$  0.15 & -1.894 $\pm$ 0.011 & 12.47 $\pm$  0.74 & -1.730 $\pm$ 0.019 & 12.0 & -1.51$^d$ & 
 0.80 \\
NGC5286 b 1 & 11.93 $\pm$  0.68 & -1.882 $\pm$ 0.014 & 12.63 $\pm$  0.79 & -1.681 $\pm$ 0.022 &  &  \\
NGC5286 c 1 & 13.13 $\pm$  0.26 & -1.827 $\pm$ 0.011 & 12.50 $\pm$  0.54 & -1.519 $\pm$ 0.011 &  &  \\
NGC5904 a 1 &  5.26 $\pm$  0.10 & -1.334 $\pm$ 0.007 &  8.07 $\pm$  0.36 & -1.247 $\pm$ 0.009 & 10.9 & -1.26$^a$ & 
 0.31 \\
NGC5904 b 1 &  5.45 $\pm$  0.09 & -1.301 $\pm$ 0.006 &  8.22 $\pm$  0.36 & -1.226 $\pm$ 0.008 &  &  \\
NGC5927 a 1 &  9.70 $\pm$  0.35 & -0.463 $\pm$ 0.011 & 11.05 $\pm$  0.49 & -0.439 $\pm$ 0.011 & 10.0 & -0.64$^b$ & 
-1.00 \\
NGC5927 b 1 & 11.08 $\pm$  0.51 & -0.512 $\pm$ 0.011 & 12.72 $\pm$  0.77 & -0.487 $\pm$ 0.011 &  &  \\
NGC5946 a 1 & 11.45 $\pm$  1.06 & -1.794 $\pm$ 0.022 & 12.33 $\pm$  1.24 & -1.568 $\pm$ 0.025 & 12.5 & -1.54$^d$ & 
 1.00 \\
NGC5986 a 1 & 15.83 $\pm$  0.11 & -1.879 $\pm$ 0.012 & 12.66 $\pm$  0.58 & -1.562 $\pm$ 0.014 & 12.0 & -1.53$^c$ & 
 0.97 \\
NGC6121 a 1 &  6.86 $\pm$  0.22 & -1.283 $\pm$ 0.012 & 10.91 $\pm$  0.80 & -1.227 $\pm$ 0.010 & 11.7 & -1.15$^a$ & 
-0.06 \\
NGC6171 a 1 &  6.60 $\pm$  0.39 & -0.974 $\pm$ 0.020 &  7.98 $\pm$  0.99 & -0.956 $\pm$ 0.014 & 11.7 & -1.13$^c$ & 
-0.73 \\
NGC6171 b 1 &  7.92 $\pm$  0.42 & -0.999 $\pm$ 0.013 & 10.58 $\pm$  0.84 & -0.974 $\pm$ 0.013 &  &  \\
NGC6218 a 1 &  7.19 $\pm$  0.23 & -1.511 $\pm$ 0.012 & 11.24 $\pm$  0.70 & -1.405 $\pm$ 0.013 & 12.5 & -1.32$^c$ & 
 0.97 \\
NGC6235 a 1 &  6.76 $\pm$  0.53 & -1.286 $\pm$ 0.032 & 11.21 $\pm$  2.34 & -1.268 $\pm$ 0.027 & 12.0 & -1.36$^c$ & 
 0.89 \\
NGC6254 a 1 & 11.74 $\pm$  0.63 & -1.770 $\pm$ 0.012 & 12.62 $\pm$  0.63 & -1.620 $\pm$ 0.016 & 11.8 & -1.51$^a$ & 
 0.98 \\
NGC6266 a 1 &  6.34 $\pm$  0.18 & -1.133 $\pm$ 0.009 &  7.99 $\pm$  0.36 & -1.112 $\pm$ 0.007 & 12.0 & -1.20$^c$ & 
 0.32 \\
NGC6284 a 1 &  5.29 $\pm$  0.14 & -1.215 $\pm$ 0.009 & 10.42 $\pm$  0.69 & -1.160 $\pm$ 0.011 & 12.0 & -1.27$^d$ & 
 1.00 \\
NGC6284 a 2 &  5.30 $\pm$  0.13 & -1.152 $\pm$ 0.008 &  8.27 $\pm$  0.46 & -1.090 $\pm$ 0.010 &  &  \\
NGC6304 a 1 & 10.12 $\pm$  0.44 & -0.580 $\pm$ 0.011 & 10.39 $\pm$  0.50 & -0.566 $\pm$ 0.011 & 10.0 & -0.66$^b$ & 
-1.00 \\
NGC6316 a 1 & 11.97 $\pm$  0.63 & -0.810 $\pm$ 0.012 & 12.51 $\pm$  0.74 & -0.799 $\pm$ 0.013 & 10.0 & -0.90$^d$ & 
-1.00 \\
NGC6316 b 1 & 10.97 $\pm$  0.53 & -0.816 $\pm$ 0.009 & 12.47 $\pm$  0.68 & -0.804 $\pm$ 0.012 &  &  \\
NGC6333 a 1 & 14.60 $\pm$  0.22 & -2.031 $\pm$ 0.013 & 14.12 $\pm$  0.26 & -1.968 $\pm$ 0.017 & 12.0 & -1.65$^d$ & 
 0.87 \\
NGC6342 a 1 &  8.96 $\pm$  0.74 & -0.910 $\pm$ 0.019 & 11.43 $\pm$  1.55 & -0.920 $\pm$ 0.018 & 12.0 & -1.01$^d$ & 
-1.00 \\
NGC6352 a 1 & 13.21 $\pm$  0.64 & -0.693 $\pm$ 0.009 & 13.25 $\pm$  0.50 & -0.693 $\pm$ 0.009 &  9.9 & -0.70$^b$ & 
-1.00 \\
NGC6356 a 1 & 11.35 $\pm$  0.41 & -0.715 $\pm$ 0.008 & 13.14 $\pm$  0.64 & -0.706 $\pm$ 0.009 & 10.0 & -0.74$^d$ & 
-1.00 \\
NGC6362 a 1 &  9.75 $\pm$  0.44 & -1.164 $\pm$ 0.014 & 13.18 $\pm$  0.65 & -1.102 $\pm$ 0.011 & 11.3 & -1.17$^c$ & 
-0.58 \\
NGC6388 a 1 &  5.77 $\pm$  0.13 & -0.590 $\pm$ 0.009 &  6.54 $\pm$  0.39 & -0.550 $\pm$ 0.008 & 10.6 & -0.68$^d$ & 
 0.00 \\
NGC6441 a 1 &  5.92 $\pm$  0.15 & -0.544 $\pm$ 0.010 &  6.88 $\pm$  0.41 & -0.522 $\pm$ 0.008 & 12.7 & -0.65$^d$ & 
 0.00 \\
NGC6441 a 2 &  5.95 $\pm$  0.15 & -0.544 $\pm$ 0.009 &  6.77 $\pm$  0.22 & -0.520 $\pm$ 0.008 &  &  \\
NGC6522 a 1 &  5.24 $\pm$  0.14 & -1.117 $\pm$ 0.008 &  8.07 $\pm$  0.49 & -1.051 $\pm$ 0.010 & 12.0 & -1.39$^c$ & 
 0.71 \\
NGC6528 a 1 & 11.10 $\pm$  0.39 & -0.279 $\pm$ 0.010 & 13.29 $\pm$  0.57 & -0.239 $\pm$ 0.009 & 10.0 & -0.10$^c$ & 
-1.00 \\
NGC6528 a 2 & 10.88 $\pm$  0.34 & -0.268 $\pm$ 0.009 & 11.98 $\pm$  0.39 & -0.230 $\pm$ 0.009 &  &  \\
NGC6528 b 1 &  8.75 $\pm$  0.29 & -0.218 $\pm$ 0.012 & 11.56 $\pm$  0.55 & -0.206 $\pm$ 0.012 &  &  \\
NGC6528 b 2 &  8.83 $\pm$  0.26 & -0.213 $\pm$ 0.011 & 11.04 $\pm$  0.43 & -0.202 $\pm$ 0.011 &  &  \\
NGC6528 c 1 & 11.35 $\pm$  0.43 & -0.306 $\pm$ 0.011 & 14.15 $\pm$  0.66 & -0.264 $\pm$ 0.010 &  &  \\
NGC6528 c 2 & 11.85 $\pm$  0.44 & -0.291 $\pm$ 0.009 & 11.85 $\pm$  0.44 & -0.291 $\pm$ 0.009 &  &  \\
NGC6544 a 1 & 13.55 $\pm$  0.32 & -1.333 $\pm$ 0.011 & 12.99 $\pm$  0.64 & -1.251 $\pm$ 0.012 & 12.7 & -1.38$^c$ & 
 1.00 \\
NGC6553 a 1 &  8.71 $\pm$  0.24 & -0.284 $\pm$ 0.010 & 14.05 $\pm$  0.66 & -0.252 $\pm$ 0.009 & 10.0 & -0.20$^c$ & 
-1.00 \\
NGC6569 a 1 & 15.00 $\pm$  0.51 & -1.028 $\pm$ 0.009 & 13.65 $\pm$  0.87 & -0.960 $\pm$ 0.013 & 10.9 & -1.08$^d$ & 
-0.51 \\
NGC6624 a 1 &  8.14 $\pm$  0.26 & -0.698 $\pm$ 0.009 &  8.21 $\pm$  0.31 & -0.692 $\pm$ 0.009 & 10.6 & -0.70$^b$ & 
-1.00 \\
NGC6624 a 2 &  8.30 $\pm$  0.26 & -0.701 $\pm$ 0.008 &  8.35 $\pm$  0.29 & -0.700 $\pm$ 0.008 &  &  \\
NGC6626 a 1 &  5.52 $\pm$  0.12 & -1.234 $\pm$ 0.007 & 12.78 $\pm$  0.53 & -1.179 $\pm$ 0.009 & 12.0 & -1.21$^c$ & 
 0.90 \\
NGC6637 a 1 & 12.96 $\pm$  0.54 & -0.857 $\pm$ 0.007 & 13.14 $\pm$  0.61 & -0.841 $\pm$ 0.007 & 10.6 & -0.78$^b$ & 
-1.00 \\
NGC6638 a 1 &  9.92 $\pm$  0.43 & -0.966 $\pm$ 0.011 & 13.14 $\pm$  0.75 & -0.936 $\pm$ 0.009 & 11.5 & -1.08$^c$ & 
-0.30 \\
NGC6652 a 1 & 10.52 $\pm$  0.36 & -0.999 $\pm$ 0.007 & 11.27 $\pm$  0.63 & -0.988 $\pm$ 0.008 & 11.4 & -1.10$^d$ & 
-1.00 \\
NGC6652 b 1 &  8.65 $\pm$  0.33 & -1.061 $\pm$ 0.007 & 11.26 $\pm$  0.73 & -1.087 $\pm$ 0.008 &  &  \\
NGC6723 a 1 &  7.29 $\pm$  0.24 & -1.299 $\pm$ 0.011 & 10.75 $\pm$  0.90 & -1.278 $\pm$ 0.011 & 11.6 & -1.14$^c$ & 
-0.08 \\
NGC6752 a 1 &  8.63 $\pm$  0.25 & -1.933 $\pm$ 0.013 & 12.60 $\pm$  0.82 & -1.864 $\pm$ 0.022 & 12.2 & -1.57$^a$ & 
 1.00 \\
NGC7078 a 1 & 10.42 $\pm$  0.45 & -2.291 $\pm$ 0.000 & 10.56 $\pm$  0.54 & -2.279 $\pm$ 0.017 & 12.0 & -2.26$^d$ & 
 0.96 \\
NGC7078 a 2 & 10.49 $\pm$  0.39 & -2.291 $\pm$ 0.000 & 10.59 $\pm$  0.47 & -2.255 $\pm$ 0.015 &  &  \\
\hline
\end{tabular}
\end{minipage}
\label{table:table1}
\end{table*}

\subsubsection{STECKMAP}

STECKMAP\footnote{
http://astro.u-strasbg.fr/Obs/GALAXIES/stecmap\_eng.html
} 
(STEllar Content and Kinematics via Maximum a Posteriori) is
a Bayesian method to interpret galaxy spectra in terms of stellar
content. It uses a non-parametric formalism, i.e. no shape is assumed
a priori for the variables, for instance the stellar age distribution.
Its principles, along with extensive simulations are
given in \citet{ocv06a,ocv06b}. The algorithm coded in
{\em{yorick}} is now reasonably user-friendly and the package
contains some plugged-in SSP models \citep{BC03,PHR,GD05}.

STECKMAP determines age and metallicity distributions (actually weights
of the SSPs of the basis).
In order to compare with the results of NBURSTS, we compute the
luminosity weighted age as:
\begin{equation}
{\rm{t_{lw}}}= 10^{\sum_{i=1}^n X_i \log({\rm{t}}_i)} \, ,
\end{equation}
where ${\rm{t}}_i$ is the age of the $i$-th SSP of the basis and 
$X_i$ is the weight (i.e. flux fraction) of this component.
Similarly, we derive the luminosity-weighted metallicity as:
\begin{equation}
{\rm{[Fe/H]_{lw}}}= 10^{\sum_{i=1}^n X_i \log([Fe/H]_i)} \, .
\end{equation}

These luminosity weighted quantities are expected to be comparable to
the SSP-equivalent values determined by other procedures, including
NBURSTS, and actually the two types of measurements are often confused
in the literature. Note however that 'luminosity-weighted' and
'SSP-equivalent' are not formally the same.

In the present paper we are interested in comparing simple populations
and will therefore not consider the details of the full stellar age
distributions recovered with STECKMAP.

\section{Analyzing one population model with another one as a reference}

In this section we will compare different models to check their 
consistency.
We will invert SSPs produced by a particular population synthesis code using 
a grid of models made with a different code, and compare the retrieved
population parameters with the nominal ones.

We restricted the comparison to intermediate-age and old populations:
600~Myr -- 15~Gyr. 
We do not discuss the younger ages because we do not have reference
spectra of
real clusters for an external validation.
Note however that we noticed a counter-intuitive dependence of the
strength of the
metallic lines in young \phr\ models. At a given age, lines like Ca at 4227~{\AA}
become stronger when the metallicity decreases, as a result of the
decreasing
temperature of the isochrones. A validation of the analysis of young
population should be carried on in a separate future work.

We also excluded the highest bin of super-solar metallicity
generated by \phr\ which is absent in BC03 and 
\vaz\ models. 

The metallicity range of the comparisons is  $-1.7 < [Fe/H] < 0.4$ 
for BC03 and  $-1.7 < [Fe/H] < +0.2$ for \vaz. For \vaz\ and BC03
we use the SSPs given by the authors. The fit was made
starting from 4000~{\AA} to be consistent with the fit
of the globular clusters (see Sec.4).

Figure~\ref{fig:1} presents the result of the inversion
of both BC03 and \vaz\ SSPs using a grid of \phr\ models. We performed
 the fit using the two inversion algorithms NBURSTS 
and STECKMAP.

The BC03 vs \phr\ comparison (Fig.~\ref{fig:1}a) reveals a reasonable 
consistency for solar metallicity, and for a wide range of ages 
(from 900~Myr to 11~Gyr).
For either sub-solar or super-solar metallicity the inversion by
\phr\ restores characteristics converging toward solar.
At metallicity of [Fe/H]~=~$-0.7$~dex the inversion is acceptable,
but degrades again in the lowest metallicity bin. 

The reasons for these discrepancies can be found in the details of
the computation of the BC03 spectra.
Indeed, the 187 stars of the BC03 library are strongly concentrated
around solar metallicity (1/3 of them have $-0.2 < [Fe/H] < 0.1$)
as in most of the other libraries. But the small number of stars remaining
at non-solar metallicity makes the modeling hazardous.

\begin{figure}
  \centering
  \subfigure(a){
  \includegraphics[width=0.49\textwidth ,height=0.3\textheight]{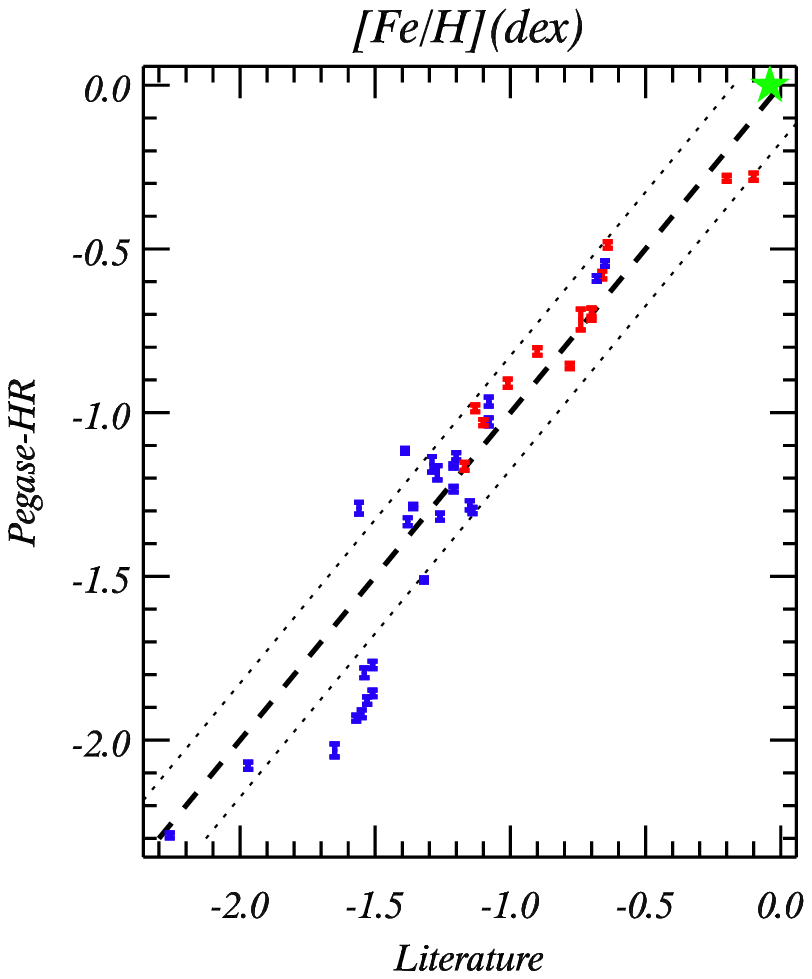}}
  \subfigure(b){
  \includegraphics[width=0.49\textwidth,height=0.3\textheight]{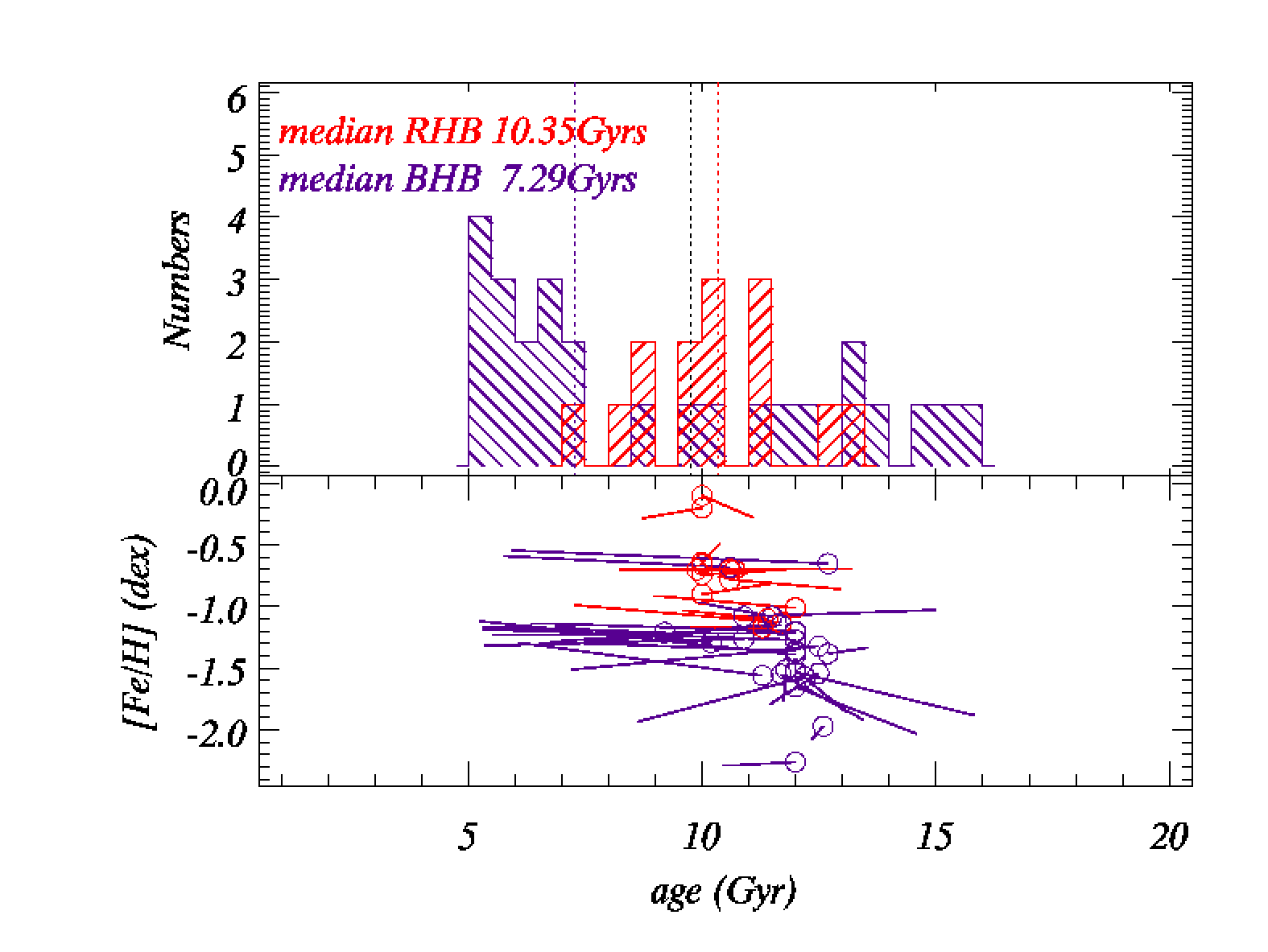}}
  \hspace{20pt}
  \caption{Relations for the GCs.
(a) - comparison  between the metallicities found with NBURSTS 
and those from \citet{schia} (spectroscopy of individual stars). 
The dotted lines represent $1~\sigma$ deviation.
 The green star represents M67;
 (b) - The upper panel shows the histogram of the ages. The dotted lines represent the
median age for the clusters;  The lower panel
compares the ages and metallicities that we find and the those from the literature.
The circles are showing the position from the literature and
the end of the vectors are our estimations.
 Everywhere blue is for the clusters with BHB and red is for the other clusters.}
  \label{fig:gc_results}
\end{figure}

\begin{figure}
  \centering
  \subfigure(a){
  \includegraphics[width=0.49\textwidth,height=0.3\textheight]{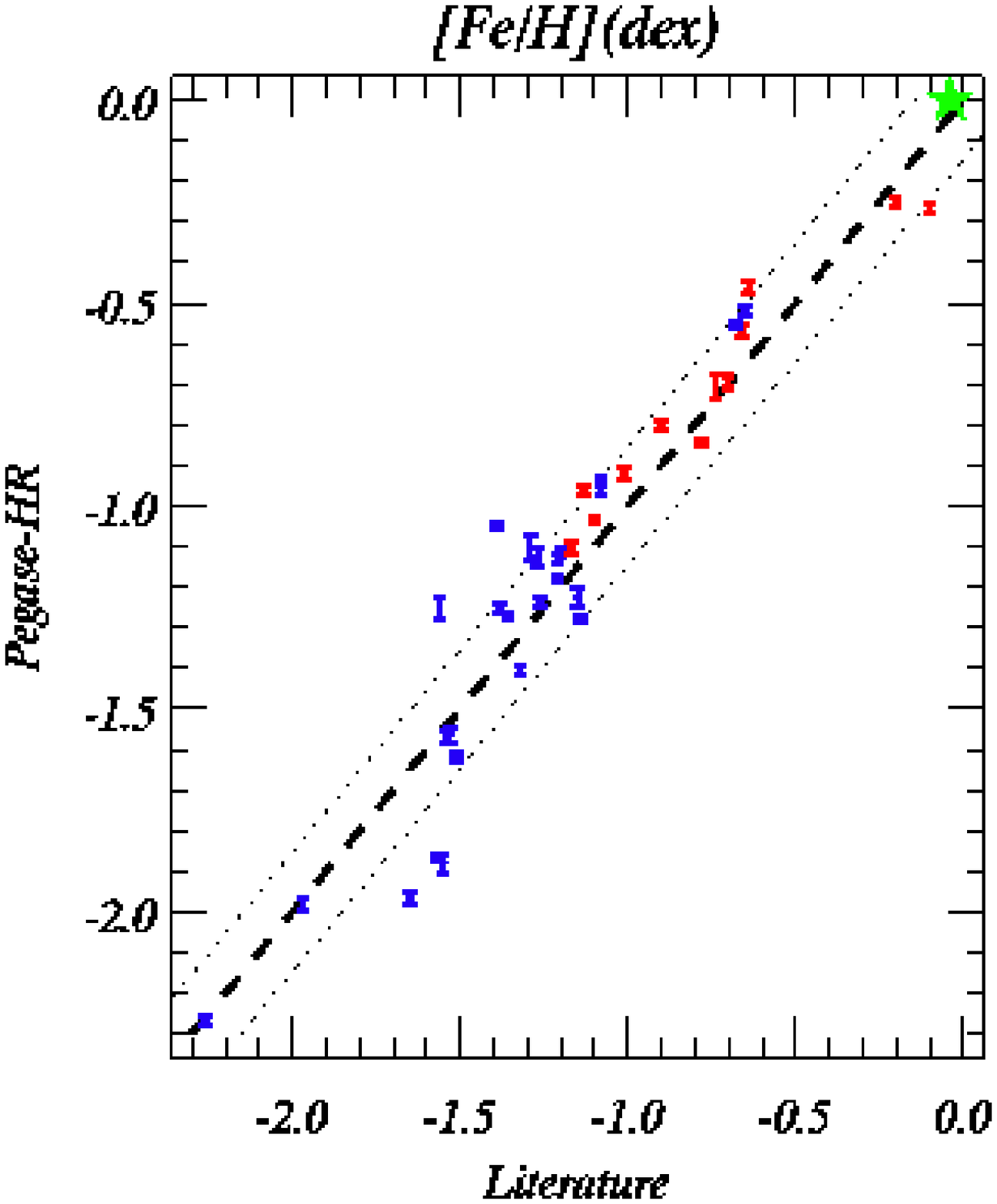}}
  \subfigure(b){
  \includegraphics[width=0.49\textwidth,height=0.3\textheight]{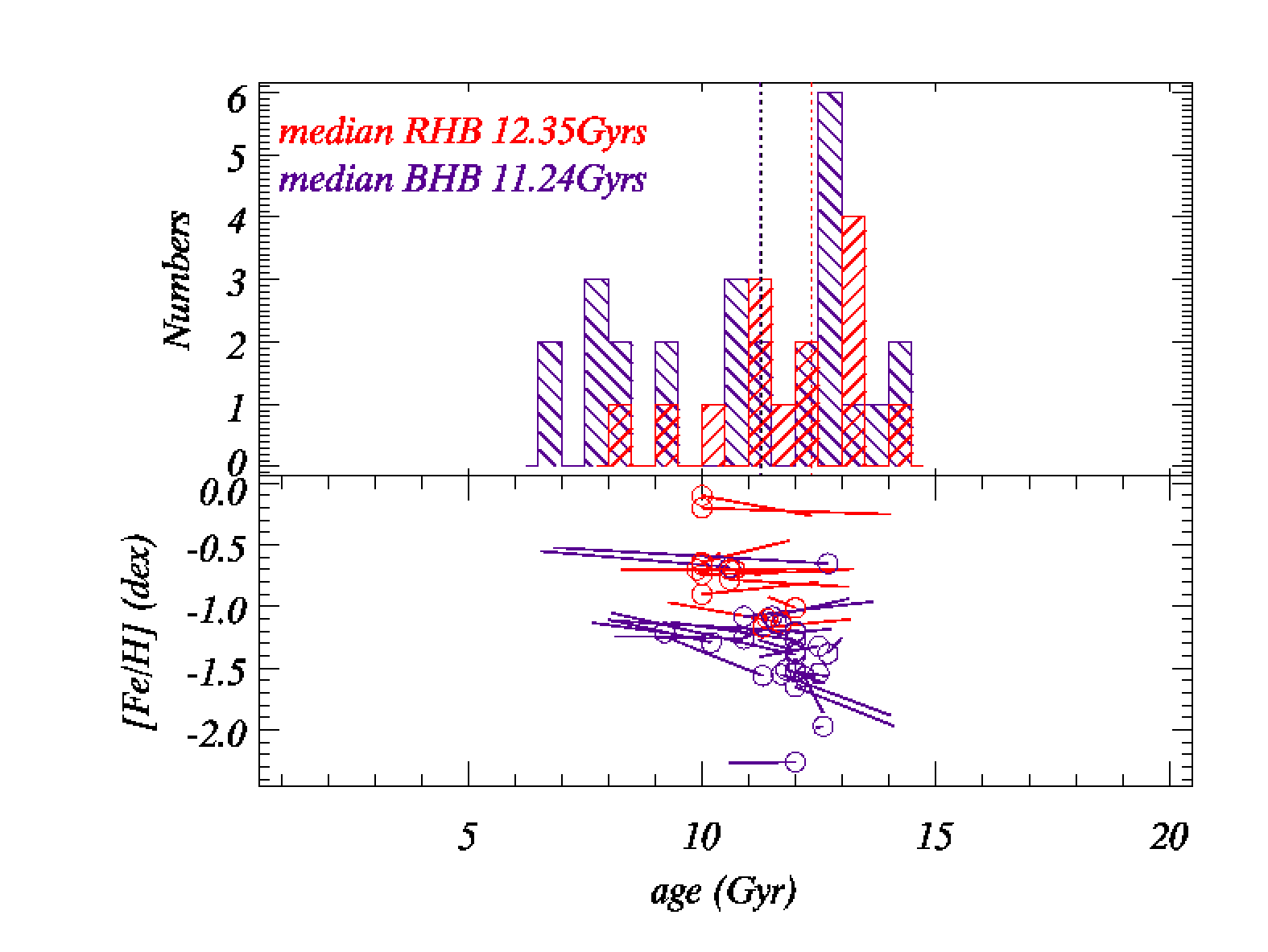}}
  \hspace{20pt}
  \caption{
Same as Fig.~\ref{fig:gc_results} but the fit was made adding hot stars
component to the SSPs models to account for the contribution
of BHB. 
}
  \label{fig:gc_nstars}
\end{figure}

To generate a SSP spectrum, BC03 uses the STELIB spectra which
are the closest to the isochrone sampling point. In particular, for each
metallicity they choose the stars having a compatible [Fe/H]. At
solar metallicity the HR diagram is sampled by 69 stars with close-to-solar 
metallicities. But for [Fe/H]~=~0.4~dex, the 41 stars that are used 
have a mean metallicity which is only slightly 
super-solar. Also, the 69 stars used for the [Fe/H]~=~$-0.4$~dex bin span a 
metallicity range from solar to [Fe/H]~=~$-0.7$~dex with an average close to 
[Fe/H]~=~$-0.2$~dex. Not surprisingly, the inversions of the corresponding 
SSPs in the age range 0.5 to 10~Gyr reproduce this convergence toward
solar metallicity.
The  [Fe/H]~=~$-0.7$~dex bin is represented by only 22 stars which are actually
well distributed around this median value, and the comparison with 
\phr\ is more consistent. The last metallicity bin of BC03,
[Fe/H]~=~$-1.7$~dex is computed from 25 stars spanning
a large range in metallicity, and the comparison with \phr\
follows the expected discrepancy.

A direct comparison between BC03 and \vaz\ gives the same results, hence 
confirming this analysis.
Another weakness of the BC03 models was revealed by our analysis. 
The inaccurate wavelength calibration of STELIB results in velocity
errors up to $40$~km\,s$^{-1}$ (shifts of $0.6$~{\AA}).
The wavelength dependence of the velocity error can be expressed as:
$ \delta v(\lambda) = 8248.5 - 6.5433\lambda + 0.001906\lambda^2 - 2.419 \times 10^{-07}\lambda^3
 + 1.128 \times 10^{-11}\lambda^4$, and the change of the velocity
dispersion:
$\sigma_{ins}(\lambda) =346.620-0.0855\lambda+0.000006\lambda^2 $, 
 $\lambda$ is in {\AA}, $\delta v$ and $\sigma_{ins}$ 
are in km\,s$^{-1}$.$\delta v$ is relative
to ELODIE.3.1 (whose wavelength calibration is
very precise), and $\sigma_{ins}$ is the absolute instrumental velocity
dispersion
deconvolved from the LSF of \phr\ ($13$~km\,s~$^{-1}$).
The LSF for the different SSPs from BC03/STELIB was found consistent
despite the inhomogeneous origins of the spectra.

Contrasting with the previous, the comparison between \vaz\ and \phr\ 
(Fig.~\ref{fig:1}b,c) is 
considerably more
consistent (the MILES library counts almost as many stars as ELODIE.3.1 and
has a comparable coverage of the space of parameters). 
\phr\ is using the Padova~94 tracks and \vaz\ Padova~2000 which,
should result in overestimating the SSP-equivalent ages 
\citep{BC03}.
The difference between the two sets of isochrones is 
compatible with the deviations between the two models on Fig.~\ref{fig:1}b,c. 

Figure~\ref{fig:vazphr} compares \phr\ and \vaz\ spectra
for three representative regions of  
Fig.~\ref{fig:1}b,c. The two models at the same age and metallicity
are over-plotted with only adjustment of the velocity dispersion and
multiplicative polynomial. The residuals are of the order of 1 per cent, even
for the lowest metallicity bin, where the S/N of the libraries is
lower because of the few stars in that region.

We also studied the dependence of the solution on the initial guesses
given to NBURSTS in order to identify cases where the solution is trapped by
a local minimum. In general, the solution is stable for a very wide range of
initial guesses and the $\chi^2$ maps 
(Fig.~\ref{fig:chi2vaz}) confirm that there is an unique global minimum.

Only in the region of old age ($>$10~Gyr) and low metallicity 
([Fe/H] $\approx$ -1.7) the topology of the  $\chi^2$ map becomes
more complex and the sensitivity to the guesses critical. 
In this regime the age dependence of most features is small and irregular.
It corresponds to the region where the 
Lick bi-indices grids, like H$_\beta$ {\it vs.} MgFe, fold on themselves, making
age and metallicity determinations ambiguous. 
This region is of importance for the study of globular clusters, and
we note that unlike Lick indices our method can determine unambiguously
the age if care is taken to avoid the secondary minimum (see Sec. 4.1).
Nevertheless, we should also stress two limitations of the models in that region
(i) the stellar libraries are only sparsely populated and (ii) the
metallicities are poorly sampled by the isochrones, the 2 values
provided in Padova~94: [Fe/H] = $-1.7$~dex and [Fe/H] = $-0.7$~dex are considerably
spaced, and intermediate models are interpolated.

To estimate the effect of the interpolation of the modeled spectra 
between the metallicities we computed spectra with intermediate metallicities
with \phr\ (performing linear interpolation) and we inverted them
with the standard \phr\ grid (NBURSTS makes spline interpolations). 
The errors are generally
negligible (0.009~dex or 2~per~cent on age, 0.006~dex on [Fe/H]), 
except at low metallicity,
i.e. between the two widely separated sets of isochrones: [Fe/H]~=~$-1.2$~dex,
where the errors are (0.09~dex or 20~per~cent in age and 0.02~dex on [Fe/H]).

\section{Analysis of Galactic clusters}
We have shown that (1) full spectrum fitting can recover SSP-equivalent 
ages and metallicities
and (2) the two recent population models are consistent 
with each 
others (BC03 has 
the disadvantage of using sparsely populated library), we want now
to assess the reliability of this analysis. For this purpose, we will analyse 
high quality spectra of Galactic clusters for which studies of deep color-magnitude 
diagrams are separately available. 

First, we will analyse the 60 spectra from the \citet{schia} library of 40 
GCs covering a wide range of metallicities. 
Second, we will fit the spectrum of the open cluster M67 from \citet{sch04}. 
The latter object will complement in the high metallicity and intermediate 
age regime the tests made on the globular clusters 
(M67 has solar metallicity and is $\approx 4$~Gyr old).

\subsection{Globular clusters}

The sample of GCs from \citet{schia} was observed at a resolution of about 3.1~{\AA}.
The observations in the range 3350 -- 6430~{\AA} 
were done with a long slit spectrograph in drift scan mode in order to
integrate the population within one core radius.
The mean S/N varies from 50 to 240~{\AA}$^{-1}$ depending on the cluster.

\subsubsection{Analysis}

The fitting was performed in the wavelength range 4000 -- 5700~{\AA}. We shortened 
the original spectra to this range because 
(i) the red part of the observations has poorer quality, due to
the correction of telluric lines and its inclusion does not improve
 the quality of the fit ($\chi^2$ is not decreased), and
(ii) when we use also
the blue region of the spectrum, including in particular the
H \& K lines, the quality of the fit is decreased and the 
results are biased by 1 Gyr toward younger ages. The S/N in
this region is lower in both the models and the observations,
but this does not explain the larger $\chi^2$ . 
The H \& K lines
are deeper in the models than in the observations. 
Figure~\ref{fig:compHK} shows that the \phr\ models are still consistent with
those of \vaz,
and the inversion of the GC observations with the
latter reproduce the
same problem.
If the mismatch comes from the observations, we can either suspect a
data-reduction problem (subtraction of diffuse light)
of a characteristics of the cluster population (Ca overabundance; but
Ca4227 is well fitted).
We did not investigate further this question and we excluded this blue
range.

We used the same grid of Pegase-HR model and multiplicative polynomial
as in the
previous section. 

The instrumental velocity dispersion measured from the calibration lamp by \citet{schia}
varies with wavelength. The instrumental velocity dispersion 
is quite constant above 5000~{\AA} at about $\sigma_{ins} = 75 - 80$~km\,s$^{-1}$ and 
rise up to over 100~km\,s$^{-1}$  at 4000~{\AA}.
Since this is a sensitive point of our method we re-determined the LSF
as explained in Sect. 3.
We found the wavelength calibration to be precise and the high order
moments insignificant,
so we only corrected the variation of $\sigma_{ins}$ using the median of the
60 individual LSFs
(one per observed spectrum).

When analysing the spectra with NBURSTS we found that at low metallicity the 
algorithm happens to stop sometimes at a local minimum.
We solved this problem starting from different guesses covering ages from 
5 to 15~Gyr and appropriate metallicities and we chose the deepest 
(hopefully absolute) minimum. We discarded solutions older 
than 15~Gyr, except for NGC~5986 for which there is no minimum younger 
than that.

At low metallicity the spectral features are not very sensitive to age
(see Sec.3, Fig.~\ref{fig:chi2vaz}). 
In the bi-index grids, this corresponds to the region where the different 
isochrones are very close and even cross each-others. In our analysis, 
this results in the local minima found for some clusters. 
Choosing carefully the initial guess (we are making 
several fits from a grid of guesses), 
it is possible to determine the real solution.

The fits for all the observations are presented in graphical form in 
Appendix A. The $\chi^2$ maps in age and metallicity allow 
to understand the degeneracies and convergence issues. At the minimum, the 
$\chi^2$ are in the range 0.63 to 2.88 and the rms residuals in the
whole wavelength range are between 0.6~per~cent and 1.9~per~cent.  
$\chi^2$ values lower than one are most probably due to overestimation of the
errors in the models.  

A close examination of the residuals reveals that the bottom of the
modeled Balmer lines are typically 1.9~per~cent deeper in the 
best-fitting models than in the observation (intensity in the center of the line). 
Forcing the fit to reduce this departure (i.e. increasing the weight of the
Balmer lines) typically leads to an increase of the age by 1~Gyr 
(i.e. 10~per~cent or 0.04~dex).
The origin of this mismatch may be either in the models or in the 
observations. We have the same residuals when fitting with the \vaz\ 
models, therefore, if it is not an observational artifact, the reason would 
have to be investigated in the physical ingredients (evolutionary tracks and IMF). 
Otherwise, an under-subtraction of the background,
could explain the residuals.

\begin{figure}
\centering
\includegraphics[width=0.49\textwidth,height=0.25\textheight]{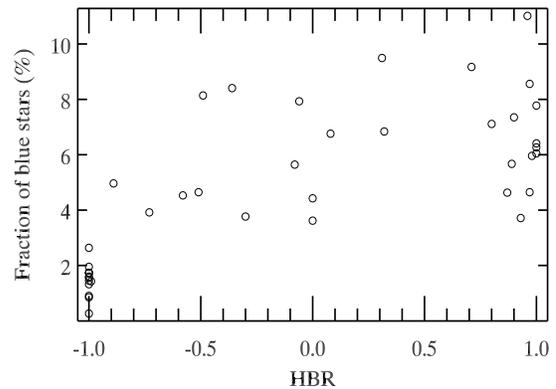}
\hspace{1pt}
\caption{Light fraction of hot stars used by the best-fitting model as a function
of the horizontal branch ratio of the clusters. As expected, the fraction
of hot stars increases with the HBR. }
\label{fig:fract}
\end{figure}

Two other features are neither fitted precisely: The Mg triplet and MgH band 
around 5175~{\AA} and the CN band near 4160~{\AA}. This is clearly an effect of 
the difference between the abundance pattern of the library and the one of the 
clusters which are $\alpha$-enhanced compared to the Sun.
At high metallicity, the library is scaled-solar and therefore clusters 
like NGC6528, NGC6553 or NGC6624 have deeper Mg than the best-fitting model.
At low metallicity, the library is also $\alpha$-enhanced (it has the 
abundance pattern of the solar neighborhood; see e. g. \citealt{wheeler89}) and 
the clusters are well fitted (rms $\approx$ 0.1~per~cent).
Figure A1  shows the effect of the abundances for different 
metallicities. Prugniel et al. (2007) have shown that using a semi-empirical 
library it is possible to improve the fit and determine the enhancement.

The CN band near 4150~{\AA} is often miss-fitted in the GCs
because non-solar abundances of C and N.
Therefore we masked the region $4145 - 4155$~{\AA} which is about half the central
bandpass of the CN1 and CN2 Lick indices.

\subsubsection{Results}

The results of the fits are presented in Table~1.
On Fig.~\ref{fig:gc_results}a we compare the value of [Fe/H] found
with our method and the one from \citet{schia}; the standard deviation 
between the two series of determinations is 0.17~dex. The reference ages are
taken mainly from \citealt{salaris02,santos04,krus06},
 for the clusters for which we did not find age reference
we adopted values of 10~Gyr.

The distribution of the ages (Fig.~\ref{fig:gc_results}b) appears more
spread than those determined from CMD analyses for which the dispersion 
is $\approx 2$~Gyr.
The effects that can alter the age determinations are (1) 
contamination of the spectrum by field stars, and (2) the horizontal branch
morphology.

Contamination by field stars may only be important in some rare cases and 
in particular for the bulge globular clusters (like for example NGC~6553).

The horizontal branch morphology or existence of blue stragglers
are well known to affect the spectroscopic
determination of the ages \citep{freitas95,lee00,maraston00}.
The horizontal branch morphology, i.e. the colour distribution of the stars 
of the horizontal branch, primarily depends on the metallicity:
The most metal poor clusters have generally the larger blue extension of the HB.
But it presents broad variations (the so-called second parameter problem,
see \citealt{rec06} for a recent dicussion)
weakly correlated with other properties of the clusters
(density, luminosity, ...). This presence
of old blue stars (Helium burning core) mimic an intermediate/young age 
population.
However, \citet{sch04b}, revisiting this question with the spectra of 
the database used in the present paper, have shown the ability to constrain 
the BHB contribution from the integrated light spectra using Fe sensitive 
line and the ratio between H$_{\beta}$ and H$_{\delta}$. 

\begin{figure*}
  \centering
  \includegraphics[width=0.9\textwidth]{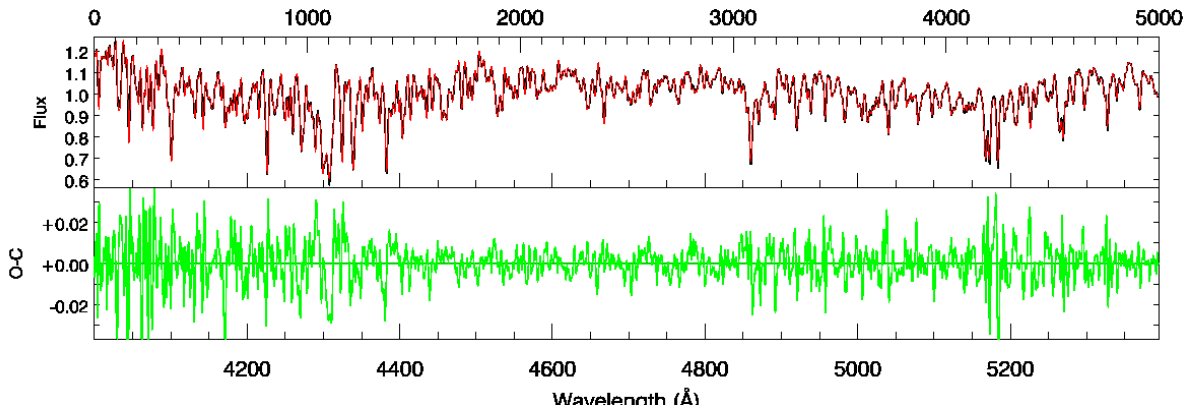}
  \hspace{20pt}
  \caption{The upper panel displays the spectrum of M67 \citep{sch04}. 
The lowed panel shows the residuals from the fit with \phr\ and NBURSTS.
Note that the vertical scale for the residuals is ten 
times expanded with respect to the plot of the spectrum.}
  \label{fig:m67}
\end{figure*}

On Fig.~\ref{fig:gc_results}  and  \ref{fig:gc_nstars} the clusters are 
separated in two groups according to the value
of the horizontal branch ratio, $HBR = (B-R)/(B+V+R)<-0.5$ \citep{harris96}.
We adopted the limit between the red and the blue HB morphology clusters 
at $HBR > -0.5$ in order to attach clusters like NGC~2808 (HBR = $-0.49$, \citealt{harris96}) 
known to have blue tail \citep{castellani06} to the 'blue' group.
Since our models do not reproduce the HB morphology, the clusters with blue HB
tend to have younger ages.

We included the effect of blue HB as an additional component in our model. 
We fitted a non-negative linear combination between the 
SSP(t, [Fe/H]) and a set 
of stars in the temperature range from 6000 to 20000~K. 
This new fit returns also
the contribution of each star in the best-fitting mix. 
These additional parameters are adding more parameters to the
fit. The consequence is that the space of parameters is more degenerated.
However, the overall chi2 values are a little bit smaller when hot stars are
added.
Figure~\ref{fig:gc_nstars}
shows that the ages of the blue clusters are moved by about 4~Gyr (0.20~dex), while the 
other clusters are less affected (0.07~dex). Only 5 (from 40) blue HB morphology clusters 
are left with an age younger than 8~Gyr. Also the metallicities are in better
agreement with the metallicities from \citet{schia}, the standard
deviation is 0.14~dex. In Fig.~\ref{fig:fract} we are showing the dependence
of the light fraction of the hot star in the best-fitting model on HBR. There 
is a tendency to have have a larger fraction of hot stars when the
HBR increases (e.g. more blue stars). 

\subsection{M67}

The Galactic open cluster, M67, is an interesting case to complete the 
comparison with the clusters since it contains an intermediate age and metal-rich
population quite similar to many diffuse elliptical galaxies 
(dE; see eg. \citealt{mich07}) or to M32.

We are using the spectrum presented in \citep{sch04} and assembled 
from spectra of individual stars in the clusters (and of some nearby
dwarfs to replace the M67 dwarfs which are too faint to be observed).
The \phr\ fit of M67 is shown in Fig.~\ref{fig:m67}.

The SSP-equivalent age $2.90 \pm 0.02$~Gyr and metallicity [Fe/H]$=-0.033 \pm 0.006$~dex are
close to those determined by spectrophotometric indices from the same 
spectrum ($ 3.5 \pm 0.5$~Gyr and [Fe/H]$=0.00$~dex, \citealt{sch04}) and 
compatible with the characteristics obtained from CMD analysis \citep{salaris02,santos04}. 

In this high-metallicity regime, the models are of the best quality as also 
shown in the comparisons of Sect. 3, and the stellar libraries sample 
ideally the corresponding
region of atmospheric parameters. For low luminosity elliptical galaxies, we
can consider our full spectrum fitting method as highly reliable.

\section{Conclusions}

We investigated the consistency and the reliability of the determination of
SSP-equivalent age and metallicity of stellar populations using full
spectrum fitting. Our method, suited to analyse medium resolution ($R > 1000$)
spectra, compares an observation to a model to derive the parameters of the
stellar population and the kinematical shifting and broadening.

We found a precise agreement between the \phr\ and \vaz\ models.
For metallicities greater than -1.0~dex and ages between 600~Myr and 15~Gyr
the metallicities of the models agree to 0.064~dex and ages to 15.5~per~cent (0.060~dex).
At low metallicity ([Fe/H]$ < -1.4$~dex) and intermediate to old ages, the age 
sensitivity is poor, and care must be taken to avoid local minima.

The BC03 models present systematic biases that are probably due to the
poor metallicity coverage of the STELIB library. The main pattern is
that they 'overestimate' the metallicity of their models for $[Fe/H]>0$~dex and they
'underestimate' the metallicity of their models for $-1<[Fe/H<0$.

Turning to real data, we found a good agreement between the SSP-equivalent
age and metallicity of M67 (2.9~Gyr and $-0.03$~dex) 
with those derived from spectrophotometric indices and from CMD analysis.
Regarding the globular clusters, the miss-match of Mg and CN 
due to non-solar abundances does not
prevent a reliable determination of [Fe/H]: The comparison
with metallicity measurements from the paper of \citet{schia}
indicate rms([Fe/H]) = 0.14~dex.

The main difficulty in the interpretation of the integrated spectra
of globular clusters, 
is the determination of the ages. In particular
the horizontal branch morphology, or presence of blue stragglers
can significantly alter the ages to misleadingly young value. The
\phr\ SSPs do not model the horizontal branch morphology and
for blue morphology the fits with NBURSTS returns ages under-estimated by
up to 30--50~per~cent. 
With STECKMAP the blue horizontal branch are identified as a 
young ($\le 300$~Myr) burst.
For the other clusters, we find a mean age of 10.4~Gyr with a dispersion 2.4~Gyr. 
Including hot stars in the NBURSTS fit to represent the blue horizontal branch,
the age of the blue clusters are restored correctly and the red clusters
are less affected. This ability to robustly measure the age of the clusters
is potentially very interesting for the study of extra-galactic clusters.

In a conclusion, as long as the population can be considered as a single 
burst, the full spectrum fitting can reliably
be used to determine ages and metallicities. It appears particularly
robust for the intermediate age populations of high metallicity
($ 1$~Gyr$ < t < 10$~Gyr, $-1 < [Fe/H] < 0$).

The full spectrum fitting uses the redundancy of the information distributed
along all the spectrum and it is not possible to identify particular
features which constrain either the age or the metallicity. Surely, the
Balmer lines are the best age indicators, but we have shown 
\citet{kol06} that the age can be retrieved even if they are masked.
In the present work each wavelength point is weighted according to its uncertainty 
and we did not intend to modulate the weight of the specific spectral features
according to their age or metallicity sensitivity (as in \citealt{pan}). 
We may in the future investigate this aspect.

An other important issue for this method is the mismatch due to
abundance differences, and in particular to $\alpha$-element enhancement.
The present population models are tied to the abundance pattern of
the solar neighborhood and cannot fit the high metallicity globular clusters
and the elliptical galaxies. But purely theoretical libraries
(see \citealt{martins07a}), or semi-empirical ones \citep{pru07b}
will soon allow to build models with variable abundances.

\section*{Acknowledgments}

The PhD grant of Mina Koleva has been provided by the French ministry of 
foreign affairs (Bourse d'excellence Eiffel doctorat). MK acknowledges 
the support of the Bulgarian National Science
Research Fund by the grant F-201/2006.
DLB thanks the Centre National d'Etudes Spatiales for its financial support.
We are grateful to I. Chilingarian, N. Bavouzet and P. Blond\'e for their
contributions to the development and tests of NBURSTS.
We thank James Rose and Ricardo Schiavon for
valuable discussions and for providing their data on the Galactic clusters.
We thank the anonymous referee, Ariane Lan\c{c}on and Eric Emsellem 
for the constructive comments.

\bsp

\label{lastpage}

\end{document}